\documentclass[11pt]{article}
\pdfoutput=1
\usepackage{hyperref}
\begin{document}
\title{Generation of Single, Monodisperse \\ Compound Droplets}
\author{James Black and G. Paul Neitzel \\
\\\vspace{6pt} George W. Woodruff School of Mechanical Engineering, \\ Georgia Institute of Technology, Atlanta, GA 30332, USA}
\maketitle
\begin{abstract}
The generation of single, monodisperse compound droplets is shown in these fluid dynamics videos.  In an apparatus designed to produce single compound droplets, a piezoelectric diaphragm generates a pressure pulse from a voltage waveform input to eject a droplet. In the method presented, oil is allowed to flow into the water nozzle with the pressure pulse ejecting both fluids as a compound droplet. Experiments were performed to demonstrate how changes in water pressure affect compound droplet compositions.  It was found that increasing the water pressure decreased the thickness of the compound droplet's oil layer.
\end{abstract}
\section{Introduction}
A compound droplet generator is used eject compound droplets.  The generator consists of a closed, pressurized water reservoir and an open silicone oil reservoir.  The pressure in the water reservoir is controlled by a pressure supply system, and the pressure in the oil reservoir is controlled by increasing its height with respect to the nozzle orifice.  

In the method presented, a small amount of oil is allowed to flow into the water nozzle and both fluids are ejected as a compound droplet. A voltage waveform drives a piezoelectric diaphragm inducing pressure waves in the generator.  The positive pressure pulse ejects a jet of fluid containing water and oil.  This fluid jet is then pinched off into a compound droplet by a negative pressure pulse. Experiments were performed to determine the affect of water pressure on the composition of compound droplets.

The videos in this submission show pressures ranging from 0.5 to 1.5 inH2O in 0.2 inH2O incremental increases.  The videos were captured using a Phantom v9.0 high speed camera attached to an Olympus SZX16 dissecting microscope.  All videos were recorded at 6400fps and played back at 8fps in the video submission.

As the pressure in the water reservoir is increased, the thickness of the oil layer is decreased.  Prior to ejection, the distance of the water-oil interface to the nozzle surface decreases as the water pressure increases.  This allows less oil to flow into the water nozzle and results in a lower percentage of oil in the ejected droplet.  In general, the pinch-off caused by the negative pressure pulse induces oscillations in the ejected droplet.  As the oil layer thickness increases, these oscillations are damped by the viscous nature of the silicone oil.  Oscillations are hardly noticeable in the low pressure tests and very apparent in the high pressure tests.
\end{document}